\documentclass[preprint,showpacs,preprintnumbers,amsmath,amssymb]{revtex4}
\usepackage{mathrsfs}
\usepackage{graphicx}
\usepackage{dcolumn}
\usepackage{bm}

\begin{document}

\title{Electron beams of cylindrically symmetric spin polarization}

\author{Yan Wang$^1$ and Chun-Fang Li$^{1,2}$\footnote{Email address: cfli@shu.edu.cn}}

\affiliation{$^1$Department of Physics, Shanghai University, 99 Shangda Road, 200444
Shanghai, China}

\affiliation{$^2$State Key Laboratory of Transient Optics and Photonics, Xi'an Institute
of Optics and Precision Mechanics of CAS, 710119 Xi'an, China}

\date{\today}

\begin{abstract}

Cylindrically symmetric electron beams in spin polarization are reported for the first
time. They are shown to be the eigen states of total angular momentum in the $z$
direction. But they are neither the eigen states of spin nor the eigen states of orbital
angular momentum in that direction.

\end{abstract}

\pacs{41.75.Fr, 41.90.+e}
\maketitle


\section{Introduction}

Both photon beams and electron beams can carry orbital angular momentum. In the case of
photon beams, the orbital angular momentum is carried by the spiral wavefront
\cite{Allen92}. Photon beams that carry orbital angular momentum have important
applications \cite{Franke} in micro-machines, atom manipulation, and quantum information.
In the case of electron beams, the orbital angular momentum was also shown
\cite{Bliokh07} to be carried by the spiral structure of the beam's wavefront when the
spin polarization is uniformly distributed. Electron beams that carry orbital angular
momentum were generated very recently in experiments \cite{Uchida, Verbeeck} by making
use of techniques that are similar to those \cite{Beij} used for generating photon beams
of orbital angular momentum.

Apart from the orbital-angular-momentum photon beams which are approximately uniformly
polarized \cite{Li09}, another kind of photon beams that are rotationally invariant in
polarization, the so-called cylindrical-vector beams \cite{Brown}, were also predicted
\cite{Hall94, Hall96} and experimentally generated \cite{Higgins, Stalder, Brown,
Kozawa}. The cylindrical-vector photon beams have drawn much attention in diverse areas
of application \cite{Zhan}. Now that the local state of quantum electron waves can be
described in terms of the concept of spin polarization \cite{Merzbacher}, an much more
interesting question arises naturally as to whether there exist electron beams that are
cylindrically symmetric in spin polarization. To the best of our knowledge, no
theoretical investigations of such electron beams have been reported.

An electron beam of cylindrically symmetric spin polarization means that it can not be an
eigen state of the spin operator in a particular direction, $\mathbf{n} \cdot
\boldsymbol{\sigma}$, where $\boldsymbol \sigma$ is the vector of Pauli matrix  and
$\mathbf n$ is a fixed unit vector. Upon noticing that the cylindrical-vector photon
beams are eigen states of the total angular momentum in the propagation direction
\cite{EnkJMO94, Li09}, denoted as the $z$ axis, our aim is therefore to look for free
electron beams that are eigen states of total angular momentum in the $z$ direction, but
not eigen states of $\sigma_z$. For this purpose, we will follow a method that is similar
to the method \cite{EnkJMO94, Li09} for constructing the cylindrical-vector photon beams.
Astonishingly, the results of our calculations reveal that so constructed free electron
beams have spin skyrmion textures \cite{Sondhi, Bogdanov, Yu}.

For simplicity only monochromatic beams are considered. The wave function in the position
representation, $\Psi(\mathbf x)$, is related to the wave function in the momentum
representation, $\psi(k_{\rho}, \varphi)$, via
\begin{equation}\label{integral}
    \Psi(\mathbf x) =\frac{1}{2 \pi} \int \psi(k_{\rho}, \varphi)
    e^{i \mathbf{k} \cdot \mathbf{x}} k_{\rho} dk_{\rho} d\varphi
\end{equation}
in circular cylindrical coordinates, where $\mathbf{k} = \mathbf{e}_x k_{\rho} \cos
\varphi +\mathbf{e}_y k_{\rho} \sin \varphi +\mathbf{e}_z k_z$ and $k_z =(k^2
-k^2_{\rho})^{1/2}$. In non-relativistic quantum frame, the Hamiltonian of free electrons
is $H =\frac{\mathbf{p}^2}{2 \mu}$ when the spin is omitted. A complete set of mutually
compatible observables in the monochromatic case can be chosen to be composed of $p_z$
and $L_z$, the $z$ components of the linear and orbital angular momenta, respectively.
The corresponding complete orthonormal set of eigen functions in the momentum
representation is given by
\begin{equation}\label{basis without spin}
    \psi_{k_z m} (k_{\rho},\varphi;\kappa) =\frac{\delta (k_{\rho} -\kappa)}
    {i^m \sqrt{2 \pi \kappa}} e^{im \varphi}, \hspace{5pt} m=0, \pm1, \pm2...
\end{equation}
where $\kappa =(k^2-k^2_z)^{1/2}$. Substituting it into Eq. (\ref{integral}) yields the
orthonormal eigen functions in the position representation,
\begin{equation*}
    \Psi_{k_z m}(\mathbf{x}; \kappa) =\sqrt{\frac{\kappa}{2 \pi}} J_m (\kappa r)
    e^{im \phi} e^{ik_z z},
\end{equation*}
where $J_m$ is the Bessel function of the first kind and $\mathbf{x} = \mathbf{e}_x r
\cos \phi +\mathbf{e}_y r \sin \phi +\mathbf{e}_z z$. Obviously, the eigen values of
$p_z$ are $k_z$ (we choose $\hbar=1$) and the eigen values of $L_z$ are $m$. They are the
scalar non-diffractive beams \cite{Durnin} in the sense that the probability distribution
does not change along with the propagation.

Equipped with the scalar basis (\ref{basis without spin}), we proceed to construct spinor
states that are cylindrically symmetric in spin polarization. In analogy with the case of
photon beams, there exist two distinct configurations. The transverse component is either
in radial direction or in azimuthal direction.

\section{Transverse component is in radial direction}

When the spin is taken into account, the free Hamiltonian is $H
=\frac{(\boldsymbol{\sigma} \cdot \mathbf{p})^2}{2 \mu}$. We first introduce a
unit-vector operator that is defined \cite{EnkJMO94, Li09} in terms of $\mathbf p$ and
the unit vector $\mathbf{e}_z$ along the $z$ axis as $\mathbf{v} =\frac{\mathbf{p} \times
\mathbf{e}_z}{|\mathbf{p} \times \mathbf{e}_z|}$. In the momentum representation, it
shows up as $\mathbf{v} =-\mathbf{e}_{\varphi}$. Then we pay attention to operator
$\boldsymbol{\sigma} \cdot \mathbf{v}$. Its normalized eigen spinors corresponding to
eigen values $\sigma=\pm 1$ have the forms of
\begin{equation}\label{eigen spinor v}
    \psi_{+1} =\frac{1}{\sqrt{2}} \left(
                                   \begin{array}{c}
                                     1 \\
                                     -ie^{i \varphi} \\
                                   \end{array}
                                 \right), \hspace{5pt}
    \psi_{-1} =\frac{1}{\sqrt{2}} \left(
                                   \begin{array}{c}
                                     -ie^{-i \varphi} \\
                                     1 \\
                                   \end{array}
                                 \right),
\end{equation}
respectively. It constitutes a complete set of mutually compatible observables together
with $p_z$ and the total angular momentum $J_z =L_z +\frac{1}{2} \sigma_z$ in the $z$
direction. Their common normalized eigen functions in the momentum representation are
proven to be
\begin{equation}\label{basis v}
    \psi_{\sigma, k_z m} (k_{\rho},\varphi;\kappa) =\psi_{\sigma}
    \psi_{k_z m} (k_{\rho},\varphi;\kappa).
\end{equation}
The eigen value of $J_z$ is $j=m +\frac{\sigma}{2}$. Substituting Eq. (\ref{basis v})
into Eq. (\ref{integral}) and replacing $m$ with $j-\frac{\sigma}{2}$, one obtains the
corresponding normalized eigen functions in the position representation,
\begin{equation}\label{ND beam v}
    \Psi_{\sigma,k_z j} (\mathbf{x}; \kappa) =\sqrt{\frac{\kappa}{4 \pi}}
    \left(
      \begin{array}{c}
        \sigma J_{j-\frac{1}{2}} (\kappa r) e^{i(j-\frac{1}{2}) \phi} \\
               J_{j+\frac{1}{2}} (\kappa r) e^{i(j+\frac{1}{2}) \phi} \\
      \end{array}
    \right) e^{ik_z z}.
\end{equation}

Let us prove that the spin polarization in these states is indeed cylindrically
symmetric. According to definition \cite{Merzbacher} $\mathbf{s} =\frac{\Psi^{\dag}
\boldsymbol{\sigma} \Psi}{\rho}$, where $\rho =\Psi^{\dag} \Psi$ is the probability
density, the transverse and longitudinal components of the spin polarization are given by
\begin{subequations}
\begin{align}
    \mathbf{s}_{\perp} & =\frac{2 \sigma J_{j-\frac{1}{2}}(\kappa r) J_{j+\frac{1}{2}}(\kappa r)}
    {J^2_{j-\frac{1}{2}}(\kappa r) +J^2_{j+\frac{1}{2}}(\kappa r)} \mathbf{e}_r, \label{T-component v} \\
    s_z                & =\frac{J^2_{j-\frac{1}{2}}(\kappa r) -J^2_{j+\frac{1}{2}}(\kappa r)}
    {J^2_{j-\frac{1}{2}}(\kappa r) +J^2_{j+\frac{1}{2}}(\kappa r)}, \label{L-component v}
\end{align}
\end{subequations}
respectively. They clearly show that (a) the transverse component is in radial direction
and both the transverse and longitudinal components are rotationally invariant,
indicating that the spin polarization is cylindrically symmetric; (b) both the transverse
and longitudinal components do not change along with the propagation, so that the beam
described by Eq. (\ref{ND beam v}) is non-diffractive in spin-polarization distribution
as well as in probability distribution.

\begin{figure}[ht]
\includegraphics[width=12.0cm]{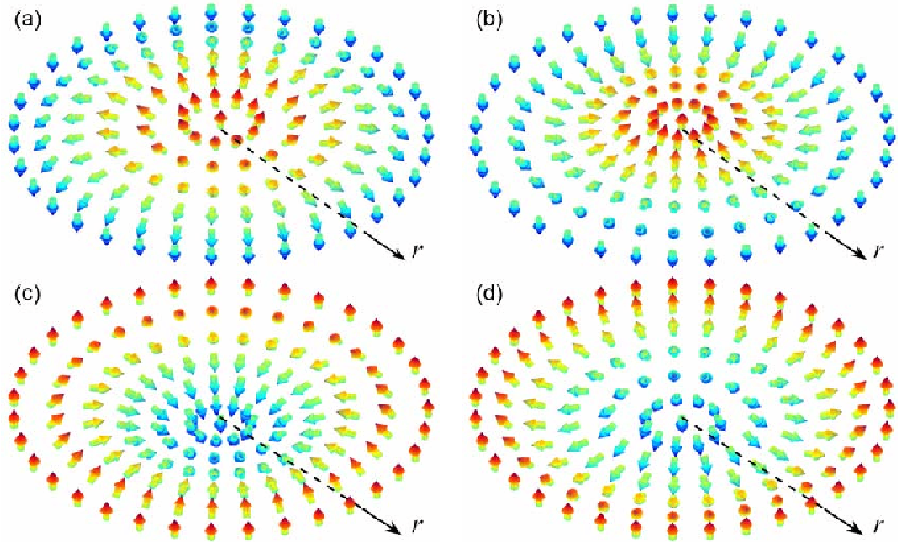}
\caption{(Color online) Schematic diagrams of the spin-polarization distributions for
non-diffractive beams of (a) $j= \frac{1}{2}$, $\sigma =1$, (b) $j= \frac{1}{2}$, $\sigma
=-1$, (c) $j= -\frac{1}{2}$, $\sigma =1$, and (d) $j= -\frac{1}{2}$, $\sigma =-1$.}
\label{nondiffraction}
\end{figure}

Eq. (\ref{T-component v}) means that the transverse component of spin polarization at
$r=0$ vanishes. That is to say, the spin polarization at the center is always
longitudinal. Based on the following asymptotic behavior of the Bessel function,
\begin{equation*}
    J_n (x) \approx \frac{x^n}{2^n n!} \hspace{5pt} \text{when} \hspace{5pt} x
    \rightarrow 0 \hspace{5pt} \text{for} \hspace{5pt} n \geq 0,
\end{equation*}
it is deduced from Eq. (\ref{L-component v}) that $s_z|_{r=0} =1$ when $j \geq
\frac{1}{2}$ and $s_z|_{r=0} =-1$ when $j \leq -\frac{1}{2}$. In addition, when either
equation $J_{j-\frac{1}{2}}(\kappa r) =0$ or equation $J_{j+\frac{1}{2}}(\kappa r) =0$ is
satisfied, the spin polarization is also purely longitudinal. Furthermore, when equation
$J^2_{j-\frac{1}{2}}(\kappa r) =J^2_{j+\frac{1}{2}}(\kappa r)$ holds, the spin
polarization is purely transverse. Consequently, as $r$ increases continuously from the
center, the spin polarization varies smoothly between transverse and longitudinal
directions alternatively, starting from a purely longitudinal one. In Fig.
\ref{nondiffraction} are schematically shown the spin-polarization distributions of
non-diffractive beams of $j =\pm \frac{1}{2}$, where the value of $\kappa r$ is taken
from $0$ to $2.4048$, the first zero point of $J_0 (\kappa r)$ at which the spin
polarization is opposite to that at the center.

Now we are ready to construct finite beams that are cylindrically symmetric in spin
polarization. The eigen functions (\ref{basis v}) or (\ref{ND beam v}) constitute an
orthonormal basis for monochromatic beams that have energy $\frac{k^2}{2 \mu}$. It is a
direct calculation to show that the following wave functions in the momentum
representation represent eigen states of $J_z$,
\begin{equation}\label{angular spectrum}
    \psi_{\sigma,m}(k_{\rho}, \varphi)
      = \int^k_0 \sqrt{\kappa} f(\kappa) \psi_{\sigma,k_z m} (k_{\rho},\varphi;\kappa)
      d\kappa,
\end{equation}
which is $\frac{1}{i^m \sqrt{2\pi}} f(k_{\rho}) \psi_{\sigma} e^{im \varphi}$, where the
scalar spectral function $f(k_{\rho})$ is square integrable and satisfies the
normalization condition $\int |f(k_{\rho})|^2 k_{\rho} dk_{\rho} =1$. As a superposition
of eigen functions $\psi_{\sigma,k_z m}$ over the eigen value $k_z$, the wave function
$\psi_{\sigma,m}$ is still an eigen state of operators $\boldsymbol{\sigma} \cdot
\mathbf{v}$ and $J_z$. Substituting Eq. (\ref{angular spectrum}) into Eq.
(\ref{integral}) and noticing the correspondence between Eqs. (\ref{basis v}) and
(\ref{ND beam v}), one arrives at the corresponding wave functions in the position
representation,
\begin{equation}\label{wave function}
    \Psi_{\sigma,j} (\mathbf{x}) =\sqrt{\frac{1}{4 \pi}}
    \left(
      \begin{array}{c}
        \sigma F_{j-\frac{1}{2}}(r,z) e^{i(j-\frac{1}{2}) \phi} \\
               F_{j+\frac{1}{2}}(r,z) e^{i(j+\frac{1}{2}) \phi} \\
      \end{array}
    \right),
\end{equation}
where
\begin{equation}\label{Fn}
    F_n (r,z) =\int^k_0 f(\kappa)J_n (\kappa r) e^{i k_z z} \kappa d\kappa.
\end{equation}
Straightforward calculations give for the transverse and longitudinal components of spin
polarization,
\begin{subequations}\label{SP}
\begin{align}
    \mathbf{s}_{\perp} & =2 \sigma
    \frac{\mathrm{Re}{(F^*_{j-\frac{1}{2}} F_{j+\frac{1}{2}})} \mathbf{e}_r
    +\mathrm{Im}{(F^*_{j-\frac{1}{2}} F_{j+\frac{1}{2}})} \mathbf{e}_{\phi}}
    {|F_{j-\frac{1}{2}}(r,z)|^2 +|F_{j+\frac{1}{2}}(r,z)|^2}, \label{T-component} \\
    s_z & =\frac{|F_{j-\frac{1}{2}}(r,z)|^2 -|F_{j+\frac{1}{2}}(r,z)|^2}
    {|F_{j-\frac{1}{2}}(r,z)|^2 +|F_{j+\frac{1}{2}}(r,z)|^2}, \label{L-component}
\end{align}
\end{subequations}
respectively. Since function $F_n$ is axially symmetric, it is clearly seen from Eqs.
(\ref{SP}) that the spin polarization is cylindrically symmetric. From Eqs. (\ref{Fn})
and (\ref{T-component}) one may deduce that the spin polarization at $r=0$ has no
transverse component. The same as before, whether it is in positive or negative $z$
direction depends on the sign of eigen value $j$. Furthermore, it can be seen from Eq.
(\ref{Fn}) that when the spectral function $f(k_{\rho})$ is a real-valued function, $F_n$
is also real-valued at plane $z=0$, the plane of beam waist. In this case, the transverse
component of spin polarization at plane $z=0$ is strictly in radial direction.

It is noted that the expectation of the spin in the state (\ref{wave function}) vanishes,
\begin{equation*}
    \int \Psi^{\dag}_{\sigma,j} (\mathbf{x}) \boldsymbol{\sigma} \Psi_{\sigma,j} (\mathbf{x})
    rdr d\phi =0.
\end{equation*}
This result can be understood as follows. Unit-vector operator $\mathbf v$ is an orbital
vector operator in the sense \cite{EnkJMO94} that it satisfies commutation relations
$[L_z,v_j] =\sum_{k} i \epsilon_{zjk} v_k$, where $\epsilon_{ijk}$ is the Levi-Civit\'{a}
pseudo tensor. So operator $\boldsymbol{\sigma} \cdot \mathbf{v}$ signifies a coupling
between the spin and orbital angular momentum. Consequently, in its eigen state
(\ref{wave function}) there exists such a coupling which is introduced through the
wave-vector dependence of the eigen spinors (\ref{eigen spinor v}). As a result of this
specific spin-orbit coupling $\boldsymbol{\sigma} \cdot \mathbf{v}$, we have the
vanishing spin angular momentum.

\begin{figure}[ht]
    \includegraphics[width=12cm]{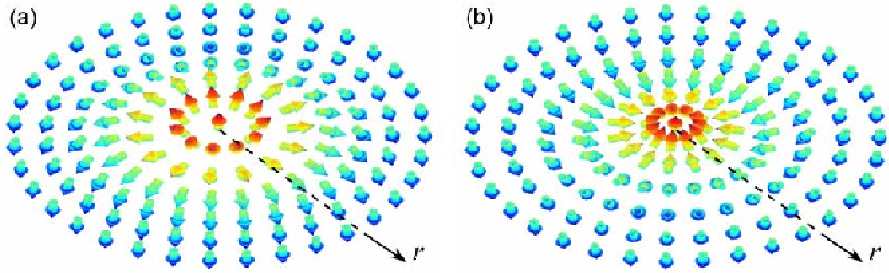}
    \caption{(Color online) Schematic diagrams of the spin-polarization distributions
for (a) $j= \frac{1}{2}$, $\sigma =1$ and (b) $j= \frac{1}{2}$, $\sigma =-1$.}
\label{skyrmion}
\end{figure}

As an example, we consider a paraxial beam which has a spectral function of Gaussian
type, $f(k_{\rho}) =\sqrt{2} w_0 \exp\left( -\frac{1}{2} w^2_0 k^2_{\rho} \right)$, where
$\frac{1}{k w_0}$ describes the divergence angle satisfying $k w_0 \gg 1$. The integral
formula
\begin{equation*}
    \int^{\infty}_0 e^{-a x^2} J_{\nu} (bx) xdx
    =\frac{\sqrt{\pi}b}{8 \sqrt{a^3}} e^{-\frac{b^2}{8a}}
     \{ I_{\frac{\nu-1}{2}} (\frac{b^2}{8a})
            -I_{\frac{\nu+1}{2}} (\frac{b^2}{8a}) \}
\end{equation*}
makes us to cope with the cases of $j\geq \frac{1}{2}$ and $j\leq -\frac{1}{2}$
separately, since it holds under the conditions \cite{Gradshteyn} of $\mathrm{Re}\nu >-2$
as well as $\mathrm{Re} a>0$. Due to the relation $J_{-n} (x) =(-1)^n J_n (x)$, we will
consider only the case of $j\geq \frac{1}{2}$ in the next of this section. Substituting
$f(k_{\rho})$ into Eq. (\ref{Fn}), making use of paraxial approximation $k_z \approx k-
\frac{\kappa^2}{2k}$ in the factor $e^{ik_z z}$, and extending the upper integral limit
to infinity, one finds
\begin{equation}\label{distribution}
    F_n(r,z)=\frac{\sqrt{\pi} w_0 r}{2 w^3} e^{-\frac{r^2}{4w^2}} \{
     I_{\frac{n-1}{2}} (\frac{r^2}{4w^2})
    -I_{\frac{n+1}{2}} (\frac{r^2}{4w^2}) \}
    e^{ikz},
\end{equation}
where $n= j\pm \frac{1}{2} \geq 0$, $w =w_0 (1+i\frac{z}{z_0} )^{1/2}$, and $z_0
=kw^2_0$. Since expression (\ref{distribution}) contains factors of Gaussian and modified
Bessel functions, the beams described by Eqs. (\ref{wave function}) and
(\ref{distribution}) are referred to as modified-Bessel-Gaussian beams. In Fig.
\ref{skyrmion} are schematically shown the spin-polarization distributions at the plane
of beam waist for (a) $j= \frac{1}{2}$, $\sigma =1$ and (b) $j= \frac{1}{2}$, $\sigma
=-1$, where $\frac{r}{w_0}$ is taken from the center to $3.36$.

An interesting phenomenon that can be seen from Fig. \ref{skyrmion} as well as Eqs.
(\ref{SP}) is that the distribution of spin polarization found here shows up as a
skyrmion texture \cite{Sondhi, Bogdanov, Yu}. The spin polarization at a cross section of
the beam is rotationally invariant about the center and varies smoothly along with the
increase of the radius. Therefore the topological charge of such a skyrmion texture
\cite{Moon}, $q= \frac{1}{2} (s_z|_{r \rightarrow \infty} -s_z|_{r=0})$, is found from
Eqs. (\ref{L-component}) and (\ref{distribution}) to be
\begin{equation*}
    q=-\frac{1}{2}\left( 1+\frac{j}{j^2+1/4} \right),
\end{equation*}
which does not change along with the beam propagation. It is dependent only on $j$, the
quantum number of $J_z$. When $j=\frac{1}{2}$, we have $q=-1$ as is clearly shown in Fig.
\ref{skyrmion}. When $j\geq \frac{3}{2}$, the topological charge is a fractional number
and approaches $-\frac{1}{2}$ in the limit $j \rightarrow +\infty$.

\section{Transverse component is in azimuthal direction}

Apart from the above mentioned beam the transverse spin polarization of which is in
radial direction at the plane of beam waist, there exists another kind of beams the
transverse spin polarization of which is in azimuthal direction at the same plane. For
the sake of completeness, we briefly summarize the essential procedures of their
construction and put forward the corresponding non-diffractive beams.

From unit-vector operator $\mathbf v$, we define another momentum-dependent unit-vector
operator $\mathbf{u} =\mathbf{v} \times \frac{\mathbf p}{p}$. The eigen spinors of
operator $\boldsymbol{\sigma} \cdot \mathbf{u}$ are as follows,
\begin{eqnarray*}
  \psi^u_{+1} &=& \frac{1}{\sqrt{2}} \left(
                                   \begin{array}{c}
                                     \sqrt{1+w_{\rho}} \\
                                     -e^{i \varphi} \sqrt{1-w_{\rho}} \\
                                   \end{array}
                                 \right), \\
  \psi^u_{-1} &=& \frac{1}{\sqrt{2}} \left(
                                   \begin{array}{c}
                                     e^{-i \varphi} \sqrt{1-w_{\rho}} \\
                                     \sqrt{1+w_{\rho}} \\
                                   \end{array}
                                 \right),
\end{eqnarray*}
which correspond to eigen values $\sigma =\pm1$, respectively, where $w_{\rho}
=\frac{k_{\rho}}{k}$. One may check that
\begin{equation}\label{basis u}
    \psi^u_{\sigma, k_z m} (k_{\rho},\varphi;\kappa) =\psi^u_{\sigma}
    \psi_{k_z m} (k_{\rho},\varphi;\kappa)
\end{equation}
is the common eigen function of operators $\boldsymbol{\sigma} \cdot \mathbf{u}$, $p_z$,
and $J_z$ in the momentum representation. The eigen value of $J_z$ is $j=m
+\frac{\sigma}{2}$. Substituting Eq. (\ref{basis u}) into Eq. (\ref{integral}) and
replacing $m$ with $j-\frac{\sigma}{2}$, one obtains the corresponding eigen functions in
the position representation,
\begin{subequations}
\begin{align}
    \Psi^u_{+1,k_z j} & =\sqrt{\frac{\kappa}{4\pi}}
    \left(
      \begin{array}{c}
          \sqrt{1+w_{\kappa}} J_{j-\frac{1}{2}} (\kappa r) e^{i(j-\frac{1}{2}) \phi} \\
        -i\sqrt{1-w_{\kappa}} J_{j+\frac{1}{2}} (\kappa r) e^{i(j+\frac{1}{2}) \phi} \\
      \end{array}
    \right) e^{i k_z z}, \nonumber \\
    \Psi^u_{-1,k_z j} & =\sqrt{\frac{\kappa}{4\pi}}
    \left(
      \begin{array}{c}
        -i\sqrt{1-w_{\kappa}} J_{j-\frac{1}{2}} (\kappa r) e^{i(j-\frac{1}{2}) \phi} \\
          \sqrt{1+w_{\kappa}} J_{j+\frac{1}{2}} (\kappa r) e^{i(j+\frac{1}{2}) \phi} \\
      \end{array}
    \right)
    e^{i k_z z}, \nonumber
\end{align}
\end{subequations}
for $\sigma =\pm1$, respectively.

The transverse and longitudinal components of spin polarization in state $\Psi^u_{+1,k_z
j}$ are given by
\begin{subequations}
\begin{align}
    \mathbf{s}^{u+}_{\perp} &
    =-\frac{2 w_z J_{j-\frac{1}{2}}(\kappa r) J_{j+\frac{1}{2}}(\kappa r) \mathbf{e}_{\phi}}
    {(1+w_{\kappa}) J^2_{j-\frac{1}{2}}(\kappa r) +(1-w_{\kappa}) J^2_{j+\frac{1}{2}}(\kappa
    r)}, \nonumber \\
    s^{u+}_z                &
    =\frac{(1+w_{\kappa}) J^2_{j-\frac{1}{2}}(\kappa r) -(1-w_{\kappa}) J^2_{j+\frac{1}{2}}(\kappa r)}
          {(1+w_{\kappa}) J^2_{j-\frac{1}{2}}(\kappa r) +(1-w_{\kappa}) J^2_{j+\frac{1}{2}}(\kappa r)},
    \nonumber
\end{align}
\end{subequations}
respectively, where $w_z =\frac{k_z}{k}$. And the transverse and longitudinal components
of spin polarization in state $\Psi^u_{-1,k_z j}$ are given by
\begin{subequations}
\begin{align}
    \mathbf{s}^{u-}_{\perp} &
    =\frac{2 w_z J_{j-\frac{1}{2}}(\kappa r) J_{j+\frac{1}{2}}(\kappa r) \mathbf{e}_{\phi}}
    {(1-w_{\kappa}) J^2_{j-\frac{1}{2}}(\kappa r) +(1+w_{\kappa}) J^2_{j+\frac{1}{2}}(\kappa
    r)}, \nonumber \\
    s^{u-}_z                &
    =\frac{(1-w_{\kappa}) J^2_{j-\frac{1}{2}}(\kappa r) -(1+w_{\kappa}) J^2_{j+\frac{1}{2}}(\kappa r)}
          {(1-w_{\kappa}) J^2_{j-\frac{1}{2}}(\kappa r) +(1+w_{\kappa}) J^2_{j+\frac{1}{2}}(\kappa r)},
    \nonumber
\end{align}
\end{subequations}
respectively. Obviously, the spin-polarization distributions in both situations are
cylindrically symmetric. But now the transverse components are all in azimuthal
direction. In much the same way as used before, one is easy to construct finite beams the
transverse spin polarization of which is in azimuthal direction at the plane of beam
waist.

\section{Conclusions and discussions}

In a word, we have found solutions of quantum electron beams that bear close analogy to
cylindrical-vector photon beams to show cylindrical symmetry in spin polarization. These
beams might be generated in experiments that are designed specially in much the similar
way to those \cite{Higgins, Stalder, Brown, Kozawa} that are designed to generate
cylindrical-vector photon beams. For example, one might consider the transmission of
uniformly polarized beams through a magnetic thin film that has a stable two-dimensional
skyrmion texture in magnetization \cite{Yu, Rossler, Yu2011}. Another possibility might
be the electron emission from a source that has a stable two-dimensional skyrmion texture
in magnetization.

CFL is indebted to Ulrich K. R\"{o}{\ss}ler for his helpful suggestions. This work was
supported in part by the National Natural Science Foundation of China (60877055 and
60806041) and the Shanghai Leading Academic Discipline Project (S30105).

\end{document}